\documentclass[reprint,aps,prl,amsmath,amssymb]{revtex4-2}

\usepackage{graphicx} % figures
\usepackage{bm}       % bold math

\setcitestyle{numbers,sort&compress}

\begin{document}

\title{Angular--Momentum--Resolved Aharonov--Bohm Coupling Energy}% Force line breaks with \\
%\thanks{Leijing}

\author{Ju Gao}
\email{jugao@illinois.edu}
\author{Fang Shen}%
\affiliation{University of Illinois, Department of Electrical and Computer Engineering, Urbana, 61801, USA}


%\affiliation{}%
%\affiliation{
%\affiliation{%

% \collaboration{CLEO Collaboration}%\noaffiliation

\date{\today}% It is today, but any date may be explicitly specified

\begin{abstract}
We present an angular--momentum--resolved energetic formulation of the Aharonov--Bohm (AB) response for a confined Dirac electron based on two gauge--invariant interaction functionals: a magnetization--field functional and a current--potential functional. Using exact Dirac eigenmodes in a cylindrical cavity threaded by a solenoidal flux, we show that the magnetization--field functional yields a core--localized interaction energy
restricted to the $l=0$ channel, with all higher angular--momentum
contributions suppressed and vanishing entirely in the limit $a\!\to\!0$. The current--potential functional, by contrast, produces a finite, mode--dependent energy shift for $l\!\ge\!1$ in the same limit, arising from a local interaction between the solenoidal vector potential and the spatially distributed Dirac current, and explicitly encoding the geometric and topological structure of the coupling energy.
\end{abstract}

\maketitle

%\tableofcontents

\section{Introduction}

The Aharonov--Bohm (AB) effect demonstrates that electromagnetic potentials can influence quantum observables even in regions where the magnetic field vanishes~\citep{Aharonov1959Significance}. 
The associated phase shift has been confirmed in numerous experiments~\citep{Chambers1960,Tonomura1986} and is widely regarded as evidence for the physical significance of the vector potential. Nevertheless, the physical origin and spatial structure of the corresponding interaction energy are less explicitly articulated in standard treatments.

Conventional descriptions of the AB effect typically emphasize global phase accumulation along a closed path~\citep{Byers1961FluxQuantization}, or invoke effective magnetic dipole couplings localized to the flux region. 
While these approaches correctly reproduce observable interference phenomena, they do not directly address how the interaction energy is distributed in space, nor which conserved physical quantities are responsible for the coupling at the level of local dynamics.

In relativistic quantum theory, an electron is described by a Dirac field supporting conserved charge and current densities. These quantities constitute the natural dynamical objects that couple to electromagnetic potentials through minimal coupling. From this perspective, the interaction between an electron and a vector potential takes the form of a local coupling between the Dirac current density and the vector potential, evaluated over the spatial extent of the electronic state.

In this work, we analyze the Aharonov--Bohm interaction energy for Dirac eigenmodes confined in a cylindrical geometry threaded by a magnetic flux. We show that a local current--potential coupling provides an energetic description that is fully consistent with the conventional phase--based formulation, while making explicit the underlying spatial structure of the interaction energy. We
further show that magnetization--field descriptions capture only a restricted contribution, corresponding to the $l=0$ angular--momentum channel of the Dirac current, whereas the full current--potential coupling resolves the mode--dependent energetic content of the flux response.

\section{Confined Dirac Modes}

We consider Dirac eigenmodes confined within a cylindrical cavity, governed by the Dirac equation with a confinement,
\begin{equation}\label{Diracwithpotential}
i\hbar \,\partial_t\Psi(\bm{r},t) =
\Big[-i\hbar c\,\bm{\alpha}\!\cdot\!\bm{\nabla}
+ \gamma^0 m_e c^2 + U(\bm{r})\Big]\Psi(\bm{r},t),
\end{equation}
where the confinement potential is taken to be
\begin{equation}\label{potential}
U(\bm{r})=\begin{cases}
0, & 0<\rho<R,\ -d<z<d,\\
U, & \rho>R,\ -d<z<d,\\
\infty, & |z|>d .
\end{cases}
\end{equation}

After separation of variables, stationary states may be written in the form
\begin{equation}
\Psi_{nlm}(\bm r,t)
= e^{-i\mathcal{E}_{nlm}t/\hbar}\,
  e^{i l \phi}\,
  \cos(k_m z)\,
  \psi(\rho),
\end{equation}
where $l=0,1,2,\dots$ is the azimuthal angular--momentum quantum number and
$k_m = m\pi/(2d)$ with odd integer $m$ enforces hard--wall confinement along the axial
direction. The radial functions $\psi(\rho)$ are spinor--valued and are determined by
boundary matching at $\rho=R$, yielding discrete radial parameters $\zeta_{nlm}$ (inside
the cavity) and $\xi_{nlm}$ (outside), together with a discrete spectrum of eigenenergies
$\mathcal{E}_{nlm}$. Such confined Dirac eigenvalue problems are well established in relativistic systems with boundaries
\citep{Chodos1974MITBag,Greiner1990RelQM} and studied for the role
of evanescent spin structure near confinement boundaries \citep{Gao2024EntropyEvaSpin}.

The full spinor solutions are presented in Ref.~\citep{GaoShen2025_JPhysComm}. For the present analysis, it suffices to note that these solutions yield explicit expressions for the charge and current densities,
\begin{equation}\label{eq:chargecurrent}
\begin{split}
q_{nlm}(\rho,z)
&= -e\,N_{nlm}^2 \cos^2(k_m z)\,\\
& \begin{cases}
J_l^2(\zeta_{nlm}\rho), & \rho<R,\\
\kappa_{nlm}^2\,K_l^2(\xi_{nlm}\rho), & \rho>R,
\end{cases}\\[3pt]
j_{nlm,\phi}(\rho,z)
&= -2\mu_B\,N_{nlm}^2 \cos^2(k_m z)\,\\
& \begin{cases}
\zeta_{nlm}\,J_l(\zeta_{nlm}\rho)\,J_{l+1}(\zeta_{nlm}\rho), & \rho<R,\\
\kappa_{nlm}^2\,\xi_{nlm}\,K_l(\xi_{nlm}\rho)\,K_{l+1}(\xi_{nlm}\rho), & \rho>R,
\end{cases}
\end{split}
\end{equation}
with $j_{nlm,\rho}=j_{nlm,z}=0$ and $\mu_B=e\hbar/(2m_e)$. The azimuthal current vanishes
on the symmetry axis as $J_l J_{l+1}\sim(\zeta\rho)^{2l+1}$, revealing a vortex--like core structure even for $l=0$, consistent with earlier analyses of spin--resolved Dirac
currents \citep{Gao2022ElectronWaveSpin,Ohanian1986}.

Normalization follows from charge conservation,
\begin{equation}\label{eq:normalizationfactor}
N_{nlm}^2
= \frac{1}{\pi R^2 d}\,
\Bigl[
\begin{aligned}[t]
- J_{l-1}(\zeta_{nlm}R)\,J_{l+1}(\zeta_{nlm}R)\\
+ \kappa_{nlm}^2 K_{l-1}(\xi_{nlm}R)\,K_{l+1}(\xi_{nlm}R)
\end{aligned}
\Bigr]^{-1}.
\end{equation}
using $\int q_{nlm}(\rho,z)\,\rho\,d\rho\,d\phi\,dz=-e$.

These confined Dirac eigenmodes render the conserved four--current fully explicit and provide the quantitative foundation for evaluating the Aharonov--Bohm coupling through distinct, gauge--invariant interaction functionals.

\section{Gauge-invariant Current--Potential Coupling}

The coupling of a Dirac electron to an external electromagnetic field is introduced by minimal coupling~\citep{BjorkenDrell1964,Sakurai1967},
\begin{equation}
H(\bm A,\phi)
= c\,\bm\alpha\!\cdot\!\big(-i\hbar\bm\nabla - q\bm A(\bm r)\big)
+ \gamma^{0} m_{e} c^{2}
+ U(\bm r)
+ q\phi(\bm r),
\end{equation}
with $q=-e$. We treat the vector potential associated with the solenoidal flux as an externally imposed field and evaluate its effect on the confined eigenmodes perturbatively. Writing $H(\bm A,\phi)=H_{0}+\hat V$, the interaction Hamiltonian
\begin{equation}
\hat V(\bm A,\phi)
= -q c\,\bm\alpha\!\cdot\!\bm A(\bm r) + q\phi(\bm r).
\end{equation}

For the solenoidal Aharonov--Bohm field we set $\phi(\bm r)=0$ and introduce the
conserved Dirac current
$\bm j(\bm r)=q c\,\Psi_{nlm}^{\dagger}(\bm r)\bm\alpha\Psi_{nlm}(\bm r)$
associated with an unperturbed eigenmode of $H_0$.
The current--potential interaction energy, defined as the first--order energy
shift, is
\begin{equation}
\Omega_{\mathrm{CP}}
\equiv \langle \Psi_{nlm}|\hat V|\Psi_{nlm}\rangle
= -\int \bm j(\bm r)\!\cdot\!\bm A(\bm r)\,d^{3}r ,
\end{equation}
which follows directly from minimal coupling (with $\phi(\bm r)=0$ here).

The functional $\Omega_{\mathrm{CP}}$ is gauge invariant under local gauge transformations
$\bm{A}\rightarrow\bm{A}+\nabla\chi$ together with $\Psi\to e^{iq\chi/\hbar}\Psi$, under which $\bm j$ is unchanged and
\begin{equation}
\delta\Omega_{\mathrm{CP}}
= -\int \bm{j}\cdot\nabla\chi\,d^3 r
= -\oint \chi\,\bm{j}\cdot d\bm{S}
  + \int \chi\,\nabla\cdot\bm{j}\,d^3 r .
\end{equation}
For stationary Dirac eigenmodes, $\nabla\cdot\bm{j}=0$, and confinement implies vanishing normal current at the boundary
($\bm j\cdot\hat{\bm n}=0$) together with evanescent decay at infinity, so the surface term vanishes for all admissible
(single-valued, nonsingular) gauge functions $\chi$, yielding $\delta\Omega_{\mathrm{CP}}=0$. The physical spectrum and transition frequencies are likewise invariant under large gauge transformations, which correspond to flux shifts by integer
multiples of $\Phi_0$ and can be absorbed by relabeling the angular--momentum index $l$.

For a solenoid aligned with the cavity axis, the vector potential takes the form
\begin{equation}\label{eq:Afield}
\bm{A}=A_{\phi}(\rho)\,\hat{\bm{\phi}},\qquad 
A_{\phi}(\rho) =
\begin{cases}
\dfrac{\Phi\,\rho}{2\pi a^{2}}, & \rho \leq a, \\[1ex]
\dfrac{\Phi}{2\pi \rho}, & \rho > a ,
\end{cases}
\end{equation}
and the magnetic field follows as
\begin{equation}
\bm{B}=B_{z}(\rho)\,\hat{\bm{z}},\qquad
B_z(\rho)=
\begin{cases}
\dfrac{\Phi}{\pi a^2}, & \rho\le a,\\[1ex]
0, & \rho>a.
\end{cases}
\end{equation}
Here $\Phi$ is the enclosed flux and $a$ regulates the solenoid core. In this configuration, the magnetic field is confined to the solenoid core, whereas the vector potential extends throughout the cavity, providing the field--free region characteristic of the Aharonov--Bohm geometry.

Substituting Eqs.~\ref{eq:chargecurrent} and \ref{eq:Afield} yields
\begin{equation}\label{eq:OmegaCP_int}
\begin{split}
\Omega_{\mathrm{CP}}
= \Omega(R)\,\Big\{
  &2\,\zeta_{nlm}\!\int_{0}^{a}\!\!
     J_l(\zeta_{nlm}\rho)\,J_{l+1}(\zeta_{nlm}\rho)\,
     \frac{\rho^{2}}{a^{2}}\,d\rho \\[3pt]
+& 2\,\zeta_{nlm}\!\int_{a}^{R}\!\!
     J_l(\zeta_{nlm}\rho)\,J_{l+1}(\zeta_{nlm}\rho)\,d\rho \\[3pt]
+& 2\,\kappa_{nlm}^{2}\,\xi_{nlm}\!\!\int_{R}^{\infty}\!\!
     K_l(\xi_{nlm}\rho)\,K_{l+1}(\xi_{nlm}\rho)\,d\rho
\Big\},
\end{split}
\end{equation}
where
\begin{equation}\label{eq:OmegaR_def}
\Omega(R) \equiv \mu_{B}\,\Phi\,d\,N_{nlm}^{2}
\end{equation}
sets the coupling scale. Using the normalization Eq.~\ref{eq:normalizationfactor}, this scale takes the explicit form
\begin{equation}\label{eq:OmegaR_explicit}
\Omega(R)
=\frac{\mu_{B}\,\Phi}{\pi R^2}\,
\Bigl[
\begin{aligned}[t]
- J_{l-1}(\zeta_{nlm}R)\,J_{l+1}(\zeta_{nlm}R)\\
+ \kappa_{nlm}^2 K_{l-1}(\xi_{nlm}R)\,K_{l+1}(\xi_{nlm}R)
\end{aligned}
\Bigr]^{-1}.
\end{equation}

Each of the three integrals appearing in Eq.~\eqref{eq:OmegaCP_int} can be reduced
using standard Bessel recurrence relations. Evaluating the resulting expressions
at the matching boundary $\rho=R$ yields
\begin{equation}\label{eq:OmegaCP}
\Omega_{\mathrm{CP}}
= \Omega(R)\,C_l(a)
+ l\,\Omega(R)\,\bigl[C_l(a)+F_l(R;a)\bigr],
\end{equation}
where
\begin{equation}\label{eq:Cla}
C_l(a)
= J_l^{2}(\zeta_{nlm}a)
- J_{l-1}(\zeta_{nlm}a)\,J_{l+1}(\zeta_{nlm}a)
\end{equation}
is a dimensionless function carrying the explicit dependence on the regulator
radius $a$, and
\begin{equation}
F_l(R;a)
= 2\left[
\int_{a}^{R}\!\frac{J_l^{2}(\zeta_{nlm}\rho)}{\rho}d\rho
+ \kappa_{nlm}^{2}\!\int_{R}^{\infty}\!
  \frac{K_l^{2}(\xi_{nlm}\rho)}{\rho}d\rho
\right],
\end{equation}
which is likewise dimensionless. For $l\ge1$, the small--core limit
$a\!\to\!0$ removes all dependence on the regulator radius, so that
$F_l(R;a)$ becomes a finite geometric function determined solely by the cavity scale $R$ and the mode parameters.

Equation~\eqref{eq:OmegaCP} makes explicit a mode--resolved interaction energy indexed by the angular momentum quantum number $l$. Its behavior in the small--core limit $a\!\to\!0$ follows directly from the Bessel expansions $J_l(x)\!\sim\!(x/2)^l/l!$ for $l\!\geq\!1$ and $J_0(x)\!\sim\!1-x^2/4$, which imply $C_l(a)\!\to\!\delta_{l0}$ for fixed $l$. In this limit the current--potential interaction energy reduces to
\begin{equation}\label{eq:OmegaCPato0}
\Omega_{\mathrm{CP}}|_{a\!\to\!0}
= \Omega(R)\,\delta_{l0}
+ l\,\Omega(R)\,F_l(R;a),
\end{equation}
showing that only the $l=0$ channel retains a contribution associated with the regulator region, while all higher-$l$
modes acquire finite, geometry-dependent energy shifts.

Equations~\eqref{eq:OmegaCP} and \eqref{eq:OmegaCPato0} therefore describe a mode--resolved interaction energy generated by local current--potential coupling, rather than by global phase accumulation. The $l$--linear contribution obtained here does not contradict the canonical Aharonov--Bohm spectrum $E_l(\Phi)\propto(l-\Phi/\Phi_0)^2$
\cite{Aharonov1959Significance,Byers1961FluxQuantization,WuYang1975PRD};
instead, it corresponds to the energetic content that gives rise to the flux dependence of that spectrum when the electronic current distribution is resolved explicitly. In effectively one--dimensional ring models~\cite{Buttiker1983PLA}, this information is absorbed into boundary conditions and thus remains implicit at the level of the spectrum. In confined Dirac systems, by contrast, the same physics appears directly through evaluation of the local interaction energy associated with the spatially extended Dirac current, thereby retaining explicit dependence on the geometry and mode structure of the electronic state that is not accessible in phase--only descriptions.

For order--of--magnitude estimates, we evaluate the characteristic scale
$\mu_{B}\Phi/(\pi R^{2})$ appearing in $\Omega(R)$.
For a cavity of radius $R=100\,\mathrm{nm}$ threaded by one flux quantum $\Phi_{0}=h/e$, this yields
$\mu_{B}\Phi/(\pi R^{2})\approx7.6~\mu\mathrm{eV}$,
corresponding to a frequency of approximately $1.8~\mathrm{GHz}$. Energy shifts of this magnitude lie within reach of modern flux--tuned cavity spectroscopy and are comparable to persistent--current splittings
observed in mesoscopic rings~\citep{Chandrasekhar1991}.
Crucially, the predicted contribution appears selectively in higher angular--momentum channels and therefore cannot be reproduced by a purely core--localized magnetization--field coupling. Spectroscopic resolution of flux--dependent level splittings across different $l$ sectors would thus provide a direct experimental signature of the spatially distributed current--potential interaction underlying the Aharonov--Bohm response.

\section{Magnetization--Field Coupling}

The angular--momentum--resolved response obtained from the current--potential interaction reflects the coupling of the solenoidal vector potential to the spatially extended Dirac current throughout the cavity. It is therefore instructive to contrast this behavior with the magnetization--field formulation, in which the electromagnetic interaction couples to a local magnetization density that is coextensive with the charge density of the confined eigenmode.
In this description, the interaction energy is determined by where the electronic density is localized, rather than by how it circulates as a current.

In the conventional quantum description, the electron is endowed with a magnetic dipole density coextensive with the
probability density, and the interaction with an
external magnetic field takes the form of a Zeeman--like energy density
$-\bm M\!\cdot\!\bm B$~\citep{Jackson1998}$,$
confined to the flux--bearing region.

For a spin--up state, the magnetization density is given by the Bohr magneton multiplied by the probability density,
\begin{equation}\label{eq:magnetization}
\begin{split}
M_{nlm,z}(\rho,z)
&= -\mu_B\,N_{nlm}^{2}\cos^{2}(k_{m} z)\,\\
&\times
\begin{cases}
J_{l}^{2}(\zeta_{nlm}\rho), & \rho \le R,\\[2pt]
\kappa_{nlm}^{2}\,K_{l}^{2}(\xi_{nlm}\rho), & \rho>R,
\end{cases}
\end{split}
\end{equation}
so that the coupling energy reads
\begin{equation}\label{eq:OmegaMF}
\begin{split}
\Omega_{\mathrm{MF}}
&= - \int \bm{M}\!\cdot\!\bm{B}\, d^3 r\\
&=-\,\frac{\Phi}{\pi a^{2}}
\int_{0}^{a}\!\!\int_{0}^{2\pi}\!\!\int_{-d}^{d}
M_{nlm,z}(\rho,z)\,\rho\,dz\,d\phi\,d\rho .
\end{split}
\end{equation}

The integrations over $z$ and $\phi$ yield factors of $d$ and $2\pi$,
respectively, leaving a core--localized radial integral,
\begin{equation}\label{eq:OmegaMF2}
\Omega_{\mathrm{MF}}
= \Omega(R)\,\frac{2}{a^{2}}
\int_{0}^{a} J_{l}^{2}\!\big(\zeta\,\rho\big)\,\rho\,d\rho. 
\end{equation}

Using the Bessel identity
\[
\frac{2}{x^{2}}\int_{0}^{x}\! J_{l}^{2}(t)\,t\,dt
= J_l^{2}(x)-J_{l-1}(x)J_{l+1}(x),
\]
the interaction energy reduces to
\begin{equation}\label{eq:OmegaMF3}
\Omega_{\mathrm{MF}} = \Omega(R)\,C_l(a).
\end{equation}

The interaction energies obtained from the two gauge--invariant functionals are thus related by the exact structural decomposition
\begin{equation}\label{eq:OmegaCP_decomp}
\Omega_{\mathrm{CP}}
= \Omega_{\mathrm{MF}}
+ l\,\Omega(R)\,\bigl[C_l(a)+F_l(R;a)\bigr].
\end{equation}

Equation~\eqref{eq:OmegaCP_decomp} shows explicitly that the
magnetization--field coupling reproduces precisely the $l=0$ contribution contained within the current--potential interaction, corresponding to a core--localized, density--weighted component of the interaction energy, while
the remaining $l$--dependent term arises from the circulating Dirac current resolved in the current--potential formulation.

In the small--core limit $a\!\to\!0$,
\begin{equation}
\Omega_{\mathrm{MF}}|_{a\!\to\!0}
= \Omega(R)\,\delta_{l0},
\end{equation}
so that only the $l=0$ channel contributes. All higher--$l$ terms vanish, in
contrast to the current--potential interaction of
Eq.~\ref{eq:OmegaCPato0}, which retains finite contributions for all angular
momenta.

\section{Discussion}

By evaluating the Aharonov--Bohm (AB) response at the level of interaction energy, we have shown that confined Dirac electrons exhibit an angular--momentum--resolved coupling that, for $l\ge1$, admits a finite, regulator--independent limit as $a\!\to\!0$ and persists even when the magnetic field is excluded from the spatial support of the Dirac current. This result complements the standard phase--based formulation of the AB effect and
provides a transparent energetic account of flux sensitivity in confined
geometries with explicitly resolved current structure.

For comparison, the original analyses of the Aharonov--Bohm effect by Aharonov and Bohm and its subsequent formulation by Byers and
Yang~\cite{Aharonov1959Significance,Byers1961FluxQuantization} showed that a confined magnetic flux modifies the electronic spectrum according to the canonical form $E_l(\Phi)\propto(l-\Phi/\Phi_0)^2$. This spectrum exhibits both quadratic and linear dependence on the angular--momentum quantum number $l$,
together with an $l$--independent flux offset.

While the conventional phase--based formulation enforces gauge invariance through global phases, the energetic Aharonov--Bohm formulation is expressed entirely in terms of local relativistic currents and electromagnetic potentials, making gauge invariance explicit and keeping the minimal--coupling structure manifest at the level of the interaction energy. In both descriptions the Aharonov--Bohm response is topological in origin; however, in the energetic formulation the topology is realized locally through
minimal coupling between current and vector potential, rather than globally through phase accumulation or boundary conditions.

As a result, the energetic formulation retains explicit dependence on the spatial structure and geometry of the Dirac current density, so that confinement shape, mode profile, and angular--momentum content enter directly into the interaction energy rather than being absorbed into a global boundary condition. From this perspective, the magnetization--field functional represents a selective $l=0$ projection of the interaction energy, whereas the
current--potential formulation preserves the full spatial and geometric content of the Aharonov--Bohm effect. Expressed at the level of interaction energy, the Aharonov--Bohm response reveals a unified structure in which topology, locality, geometry, and gauge invariance appear simultaneously.

\section{Conclusion}

The present analysis treats a confined Dirac electron as a spatially distributed
current that couples locally to the solenoidal vector potential throughout the
cavity. Within this description, the flux dependence of the spectrum emerges
from a mode--resolved interaction energy, with a finite, core--independent
contribution surviving in the small--core limit $a\!\to\!0$.

The resulting angular--momentum--dependent energy shift is experimentally
accessible through high--resolution cavity spectroscopy or flux--tuned level
splitting. For cavity radii $R\!\sim\!100~\mathrm{nm}$, the characteristic scale
$\mu_{B}\Phi/(\pi R^2)$ lies in the microwave regime, comparable to
persistent--current splittings observed in mesoscopic rings. Observation of
flux--dependent level splittings across higher angular--momentum channels would
therefore provide direct evidence of a local coupling between the solenoidal
vector potential and the spatially distributed Dirac current.

Taken together, these results show that an energetic formulation of the
Aharonov--Bohm effect captures more than the global phase shift encoded in
boundary conditions. By expressing the response directly through an observable
interaction energy, the present approach renders the interplay of topology,
locality, geometry, and gauge invariance explicit, while retaining direct
sensitivity to the spatial structure and geometry of the electronic mode.

% \section*{\label{sec:Acknowledgement}Acknowledgement}

%\appendixpage
%%%%%%%%%%%%%%%%%%%%%%%%%%%%%%%%%%%%%%%%%%%%%%%%%%%%%%%%%%%%%%%%%%%%%%%%%%%%
% Appendix: AB-like Energy Shift as Proposition–Proof–Corollary
%%%%%%%%%%%%%%%%%%%%%%%%%%%%%%%%%%%%%%%%%%%%%%%%%%%%%%%%%%%%%%%%%%%%%%%%%%%%

%\appendix

\bibliographystyle{apsrev4-2}

\bibliography{spin}  % uses Spin.bib

\end{document}